\documentclass[english,aps,twocolumn,superscriptaddress,groupedaddress,nofootinbib]{revtex4}
\usepackage{amsmath,amssymb}
\usepackage{graphicx}
\usepackage{epsfig,color}
\usepackage{bm}
\usepackage{verbatim}
\usepackage{ulem}

\begin{document}

\title{The new $P_{\rm cs}(4459)$, $Z_{\rm cs}(3985)$, $Z_{\rm cs}(4000)$ and $Z_{\rm cs}(4220)$ and the possible emergence of flavor pentaquark octets and tetraquark nonets}

\author{J. Ferretti}
\affiliation{Department of Physics, University of Jyv\"askyl\"a, P.O. Box 35 (YFL), 40014 Jyv\"askyl\"a, Finland}
\author{E. Santopinto}
\email{santopinto@ge.infn.it}
\affiliation{INFN, Sezione di Genova, Via Dodecaneso 33, 16146 Genova, Italy}

\maketitle

Recently, the LHCb and BESIII collaborations provided results for two $Z_{\rm cs}$ tetraquarks and a $P_{\rm cs}$ pentaquark. 
These are hidden-charm four- and five-quark states, characterized by $c \bar c s\bar n$ ($c \bar c n\bar s$) and $c \bar c nns$ quark structures, respectively, where the notation $n$ indicates a $u$ or $d$ quark.
They are the strange counterparts of the $Z_{\rm c}$ and $P_{\rm c}$ exotics, which are of the $c \bar c n\bar n$ and $c \bar c n n n$ type.
The $Z_{\rm cs}(3985)$ and $Z_{\rm cs}(4000)$ tetraquarks were discovered by BESIII and LHCb, respectively and the $P_{\rm cs}(4459)$ was observed by LHCb.
For more details, see the supplemental material.
The experimental finding of the $Z_{\rm cs}$ and $P_{\rm cs}$ states was anticipated in some theoretical models, including the compact tetraquark and pentaquark scenarios, the $D_{\rm s} \bar D^*$ and $\Xi_{\rm c}^0 \bar D^{*0}$ molecular models, and the hadro-charmonium model.

The experimental observation of these exotic hadrons, and the possibility of filling SU(3) flavor tetraquark nonets and pentaquark octets, might help us to discriminate among the different interpretations available for tetraquarks and pentaquarks.
Indeed, open-strange states and their complete SU(3)$_{\rm f}$ multiplets are predicted in the compact tetra/pentaquark scenarios on the basis of the SU(3)$_{\rm f}$ flavor symmetry; see e.g. Refs. \cite{Anwar:2018sol,Maiani:2021tri,Ferretti:2020ewe}.
If the observed exotics are molecules, one may not necessarily expect to find complete SU(3)$_{\rm f}$ multiplets \cite{Santopinto:2016pkp}, although this might depend on the specific choice of the molecular model parameters. Similarly, in the hadro-charmonium model the emergence of full SU(3)$_{\rm f}$ multiplets depends on the values of the parameters\footnote{Besides the problem that the charmonium chromo-electric polarizability is a poorly known parameter, the attractiveness of the hadro-charmonium potential also depends on the inverse cube of the size of the light hadron \cite{Ferretti:2020ewe}.}. 
Those models (hadro-charmonium and molecular), which do not necessarily require complete filling of the multiplets, may be favored if not all the multiplet members are observed by the experimentalists.
On the other hand, it should be borne in mind that the three above interpretations might just be different approximations of the same complex configuration.
The Born-Oppenheimer approach might thus help us to gain further insight into the properties of these multiquark states \cite{Braaten:2014qka}.

In the following, we discuss the hadro-charmonium, molecular and compact tetraquark model descriptions of hidden-charm tetraquark multiplets,  and specifically of the nonet(s) characterized by $1^{+}$ quantum numbers. 
We also discuss the classification of the ground-state $P_{\rm c}$, $P_{\rm cs}$ and $P_{\rm css}$ pentaquarks in terms of SU(3)$_{\rm f}$ octet structure(s) within the hadro-charmonium, compact pentaquark and molecular models.

{\it $X_{\rm c}$, $Z_{\rm c}$ and $Z_{\rm cs}$ states and their nonet structure in the hadro-charmonium model.} 
It is interesting to try to accommodate the existing experimental data, which are compatible with $1^+$ quantum numbers, within the hadro-charmonium multiplet in Fig. \ref{fig:nonet}. For the values of the hadro-charmonium masses and the parameter values, see the supplemental material and Ref. \cite{Ferretti:2020ewe}.
%%%%%%%%%%%%%%%%%%%%%%%%%%%%%%%%%%%%%%%%%%%%%%%%%%
\begin{figure*}[htbp] 
\centering 
\includegraphics[width=15cm]{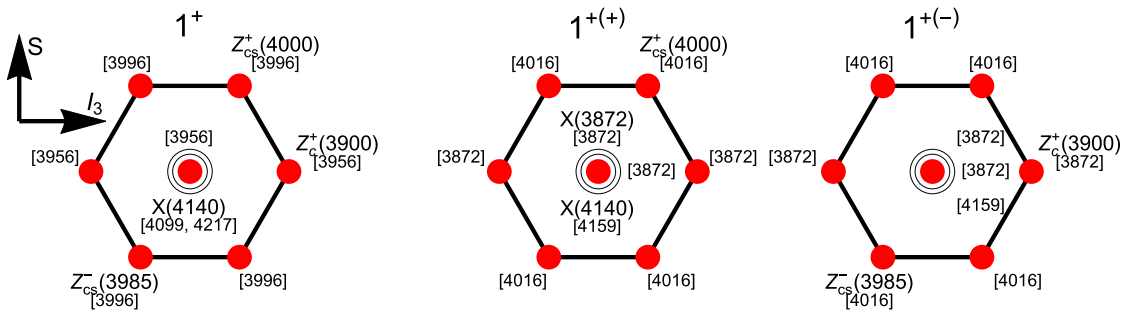}
\caption{The $1^{+}$ tetraquark nonet in the hadro-charmonium model ({\bf left}); the $1^{+(+)}$ and $1^{+(-)}$ tetraquark nonets in the compact tetraquark model ({\bf centre and right}). 
The numbers in the square brackets are the hadro-charmonium/compact tetraquark model predictions; see the text and Refs. \cite{Ferretti:2020ewe,Anwar:2018sol} for more details. The experimental masses are in round brackets. 
The tetraquark states are classified in terms of strangeness, $S$, and $I_3$ quantum numbers. The $C$-parity of the second and third multiplets refers to the sign of charge conjugation of the neutral-non-strange members.} 
\label{fig:nonet}
\end{figure*}
%%%%%%%%%%%%%%%%%%%%%%%%%%%%%%%%%%%%%%%%%%%%%%%%%%

Let us start from the $Z_{\rm cs}(3985)^-$ and $Z_{\rm cs}(4000)^+$; these can be assigned to $\psi(2S) \otimes K^-$ and $\psi(2S) \otimes K^+$ hadro-charmonia, respectively, the difference between the calculated masses \cite{Ferretti:2020ewe} and the experimental ones being less than 10 MeV. If these assignments are correct, the two $1^+$ neutral states, $\psi(2S) \otimes K^0$ and $\psi(2S) \otimes \bar K^0$, will soon be found.
It is noteworthy that this particular assignment would require the $Z_{\rm cs}(3985)^-$ and $Z_{\rm cs}(4000)^+$ to be charge conjugates; however, the large difference (not compatible with the experimental errors) in their decay widths would suggest that they should be accommodated in two different multiplets.
While this is not possible in the hadro-charmonium model alone, it can be done in the compact tetraquark scenario, as discussed in the following.
A second possibility would be to interpret the $Z_{\rm cs}(3985)^-$ as a hadro-charmonium and the $Z_{\rm cs}(4000)^+$ as a different kind of exotic; see e.g. the molecular model interpretation from Ref. \cite{Meng:2021rdg}.
One may try to assign the $\psi(2S) \otimes \pi^+$ tetraquark to the $Z_{\rm c}(3900)^+$. The large difference between the experimental and calculated masses, $\mathcal O$(70 MeV), can be explained in terms of the computational difficulties related to the smallness of the pion mass. 
One of the $\psi(2S) \otimes \eta$ and $\psi(2S) \otimes \eta'$ states, or even a superposition of them, could be assigned to the $X(4140)$.
In this picture, one cannot accommodate the $X(3872)$ [also named $\chi_{\rm c1}(3872)$ on the PDG] in the multiplet of Fig. \ref{fig:nonet}.
In this respect, it should be observed that the decay patterns of the $X(3872)$ seem to favor the interpretation of the $X(3872)$ as a $\chi_{\rm c1}(2P)$ charmonium state plus meson-meson components, like $D \bar D^*$ and so on. 

Finally, one may also consider the hypothesis of the emergence of a meson octet (plus a singlet), instead of a nonet, in the hadro-charmonium sector. This would happen if the strange and non-strange components in the $\eta$ and $\eta'$ mesons were not decoupled.

{\it$X_{\rm c}$, $Z_{\rm c}$ and $Z_{\rm cs}$ states and their nonet structure in the compact tetraquark model.} 
In order to accommodate the observed hidden-charm tetraquarks in the compact tetraquark scenario, one may consider various sets of predictions of their masses.
Here, we have chosen to use the results of Refs. \cite{Ferretti:2020ewe,Anwar:2018sol}.
The multiplets built upon them are shown in Fig. \ref{fig:nonet}.
A similar qualitative description was suggested in Ref. \cite{Maiani:2021tri}.

Unlike the hadro-charmonium case, in which there is only one $1^+$ nonet, in the compact tetraquark scenario there are three ground-state multiplets with $J^P = 1^+$: these will be denoted as $1^{+(+)}$, $1^{+(-)}$ and $1^{+(-)}_2$ (for the construction of the compact tetraquark states, see e.g. \cite[Appendix]{Anwar:2018sol}). 
The difference between the $1^{+(-)}$ and $1^{+(-)}_2$ multiplets is that the states belonging to the first are the combination of a scalar and an axial-vector diquark/antidiquark, while those belonging to the $1^{+(-)}_2$ nonet are the combination of two axial-vector diquark/antidiquark.

It should also be borne in mind mind that, in addition to the above $1^{+(+)}$,$1^{+(-)}$ and $1^{+(-)}_2$ multiplets, one also has two lower-lying (i.e. $S$-wave) $0^{+(+)}$ nonets and one $2^{+(+)}$ nonet.

The $X(3872)$, the $X(4140)$ and the $Z_{\rm cs}(4000)^+$ can be assigned to the $1^{+(+)}$ nonet.
The $Z_{\rm c}(3900)$ and the $Z_{\rm cs}(3985)$ can be assigned to the $1^{+(-)}$ nonet.
This scenario would explain why the $Z_{\rm cs}(3985)^-$ and $Z_{\rm cs}(4000)^+$ have very different decay widths (there is an order of magnitude between them).
It is worth noting that the tetraquark multiplet structure may allow for a certain amount of mixing between different multiplets in the strange sector, in which C-parity is not a well-defined quantum number.
For instance, assuming that the $Z_{\rm cs}(3985)$ and $Z_{\rm cs}(4000)$ actually belong to multiplets in which the electrically neutral members have quantum numbers $1^{+-}$ and $1^{++}$, then these strange components both have $J^P = 1^+$ and might very well show a certain degree of admixture. This would be analogous to the difference between the $K_{1A}$ and $K_{1B}$ states and the $K_1(1270)$ and $K_1(1400)$ states in the light sector.
In order to obtain a clearer picture of the properties of these mesons, their strong decay widths, which were calculated in the molecular interpretation in Ref. \cite{Meng:2021rdg}, also need to be computed within the compact tetraquark and hadro-charmonium models.

In the compact tetraquark picture, the $X(3872)$ is part of a tetraquark multiplet.
If this is the case, then: either the $X(3872)$ is a superposition of a four-quark and a charmonium $c \bar c$ component; or a new charmonium state with $1^{++}$ quantum numbers and a mass of around 3.9 GeV has to be found by the experimentalists, in order to complete the $\chi_{\rm c}(2P)$ spin charmonium-multiplet.
However, we must remember that: a) the wave function of an exotic hadron (which is directly related to its nature as a compact state, hadro-charmonium, molecule ...) is not an observable. It can sometimes be inferred to a certain extent from the decay patterns of the hadron and, in particular, from the magnitude of its total decay width. In other cases the situation is more blurred and it is therefore difficult to draw a clear conclusion; b) the experimental searches are far from complete and it is unlikely that we will end up with full multiplets of exotics in the near future. For the $X(3872)$ interpretation, see also Ref. \cite{Guo:2014taa}.

The $Z_{\rm c}(4020)^\pm$ has a mass of $4024.1\pm1.9$ MeV and its quantum numbers are expected to be $J^{P} = 1^{+}$. It can be assigned to the $1^{+(-)}_2$ multiplet \cite{Maiani:2021tri,Anwar:2018sol} and, in the relativized diquark model \cite[Table III]{Anwar:2018sol}, its calculated mass is 4047 MeV. The $1^{+(-)}_2$ multiplet also includes: two more $Z_{\rm c}$-type states and a single $X$-type state, all with masses of 4047 MeV; a single $X$-type state, with a mass of 4074 MeV and a $s \bar s c \bar c$ quark structure; and four $Z_{\rm cs}$-type states, with a mass of 4061 MeV and a $c s \bar c \bar n$ ($c n \bar c \bar s$) quark structure \cite{Anwar:2018sol,Ferretti:2020ewe}. 

It is also worth commenting on the $Z_{\rm cs}(4220)$ meson observed by LHCb, with significance exceeding five standard deviations. Its mass, width and favored quantum numbers are: $M = 4216\pm24^{+43}_{-30}$ MeV, $\Gamma = 233\pm52^{+97}_{-73}$ MeV and $J^P = 1^+$ or $1^-$, with a $2\sigma$ difference in favor of the first hypothesis.
In the compact tetraquark model, on the basis of mass predictions, one would assign it to the first $1^{-(-)}$ state of \cite[Table 7]{Ferretti:2020ewe}, with a mass of 4234 MeV.
In the hadro-charmonium picture \cite[Table 4]{Ferretti:2020ewe}, one would assign the $Z_{\rm cs}(4220)$ to a $\psi(2S)\otimes K^*$ or $\chi_{\rm c1}(2P)\otimes K^*$ state. The former has a mass of 4207 MeV and $J^P = 0^+, 1^+$ or $2^+$ quantum numbers; the latter has a mass of 4250 MeV and $J^P = 0^-, 1^-$ or $2^-$ quantum numbers. See also Ref. \cite{Voloshin:2019ilw}, in which Voloshin predicted the emergence of hadro-charmonium states with quantum numbers $J^P = 0^+$ and $1^+$ at energies around 4250 MeV and 4350 MeV, respectively.
Because of the large width of the $Z_{\rm cs}(4220)$, the molecular model interpretation of this state is unlikely.
However, in Ref. \cite{Meng:2021rdg} it was argued that the $Z_{\rm cs}(4220)$ might be interpreted as a $\bar D_{\rm s}^* D^*$ resonance with $J^P = 1^+$, which could be the heavy-quark spin symmetry (HQSS) of the $Z_{\rm cs}(4000)$.

Finally it is worth commenting on why the states belonging to the multiplets in Figs. \ref{fig:nonet} are characterized by widths which vary in a relatively wide range. 
The reason why the widths of the $Z_{\rm cs}(3985)$ and $Z_{\rm cs}(4000)$ are so different could be related to their different nature or quark structure.  
What we know for sure is that the above states cannot be part of the same multiplet because, if they were, then they would be charge conjugates, like $K^+$ and $K^-$, and have the same properties (like mass or decay width) within the experimental error.
The $X(3872)$ is extremely close to the $D\bar D^*$ threshold. This may be why its properties are distinct from those of the other members of its multiplet. 
The X(4140) might be the $c\bar c s\bar s$ counterpart of the $X(3872)$ and thus have similar properties, even though the fact that it is less close to a meson-meson threshold would make its width relatively larger.
It is also important to remember that the SU(3)$_{\rm f}$ symmetry is broken. Therefore, the states belonging to the same SU(3)$_{\rm f}$ multiplet do not necessarily have to show the same features. 

{\it$X_{\rm c}$, $Z_{\rm c}$ and $Z_{\rm cs}$ states: their spin and SU(3) flavor structures in the molecular model.}
Firstly, it should be noted that the $Z_{\rm cs}(3985)$ and $Z_{\rm cs}(4000)$, whose widths are $12.8^{+5.3}_{-4.4}\pm3.0$ MeV and $131\pm15\pm26$ MeV, respectively, are unlikely to be two different states of charge of the same meson, like $K^+$ and $K^-$ in the case of the kaon. Moreover, the $Z_{\rm cs}(4000)$ might be too large to be interpreted as a molecule, even though Ref. \cite{Meng:2021rdg} explores this possibility (plus the fact that $Z_{\rm cs}(4000)$ and $Z_{\rm c}(3900)$ are SU(3) flavor partners).
A second possibility, studied in Ref. \cite{Du:2022jjv}, is that the $Z_{\rm cs}(3985)$ and $Z_{\rm c}(3900)$ are SU(3) flavor partners (i.e. they are members of the same $J^P = 1^+$ flavor octet), the first state being described as a $(D^{*0} \bar D_{\rm s} + D^0 \bar D_{\rm s}^*)$ and the second as a $D \bar D^*$ molecule. 

Besides $Z_{\rm c}$s and $Z_{\rm cs}$s, there are also molecular model assignments for $X_{\rm c}$-type states, such as the $X(3872)$ and $X(4140)$. Owing to its proximity to the $D^0 \bar D^{*0}$ threshold, the $X(3872)$ can be interpreted as a $D \bar D^*$ meson-meson molecule with $1^{++}$ quantum numbers; see e.g. Ref. \cite{Hanhart:2007yq}.
Similarly, the $X(4140)$ can be described as a $D_{\rm s}^{*+} \bar D_{\rm s}^{*-}$ molecule \cite{Liu:2009ei}.

The molecular model also suggests that some of the observed $Z_{\rm c}$ states could be spin partners.
For example, consider the mass difference between the $Z_{\rm c}(3900)$ and $Z_{\rm c}(4020)$, which is very similar to that between the $D$ and $D^*$ mesons. This is one of the arguments in favor of their molecular nature; the HQSS also predicts the same molecular potential for the $J^{PC} = 1^{+-}$ $D^* \bar D$ and $D^* \bar D^*$ molecules. The same pattern seems to be repeated in the hidden-bottom sector, see e.g. Ref. \cite{Bondar:2011ev}, specifically in the $Z_{\rm b}(10610)$ and $Z_{\rm b}(10650)$ cases.

Predictions for the HQSS partners of the $Z_{\rm c}(3900)$ and $Z_{\rm cs}(3985)$, dubbed $Z_{\rm c}^*$ and $Z_{\rm cs}^*$, were given in \cite[Table II]{Du:2022jjv}. In this particular fit, one obtains the results: $M[Z_{\rm c}^*] = 4041^{+4}_{-5}$ MeV and $M[Z_{\rm cs}^*] = 4142^{+4}_{-5}$ MeV.

%%%%%%%%%%%%%%%%%%%%%%%%%%%%%%%%%%%%%%%%%%%%%%%%%%
\begin{figure*}[htbp]
\centering
\includegraphics[width=10cm]{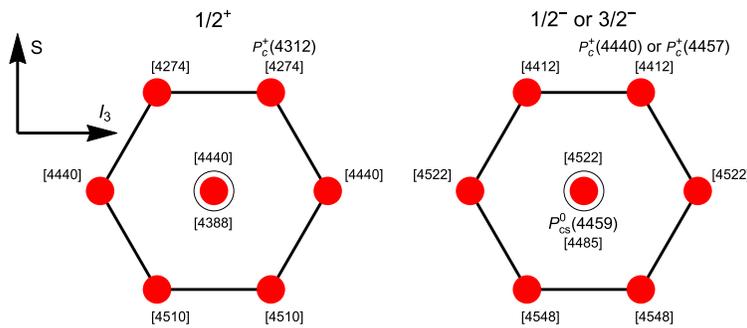} 
\caption{$J^P = \frac{1}{2}^+$ pentaquark octet, made up of the bound states of octet baryons plus the $\chi_{\rm c0}(1P)$ (left), and $J^P = \frac{1}{2}^-$ or $J^P = \frac{3}{2}^-$ pentaquark octet, made up of the bound states of octet baryons plus the $\psi(2S)$ (right). The predicted hadro-charmonium masses are reported inside square brackets, the experimental ones are inside round brackets. See the supplemental material for more details.} 
\label{fig:octet}
\end{figure*}
%%%%%%%%%%%%%%%%%%%%%%%%%%%%%%%%%%%%%%%%%%%%%%%%%%

{\it $P_{\rm c}$, $P_{\rm cs}$ and $P_{\rm css}$ pentaquarks and their octet structure.} 
In the hadro-charmonium model, $P_{\rm c}$, $P_{\rm cs}$ and $P_{\rm css}$ pentaquarks are described as the bound states of charmonia and light baryons. For further details, see the supplemental material.
The most likely assignments to the lowest $J^P = \frac{1}{2}^+$, $\frac{1}{2}^-$ and $\frac{3}{2}^-$ hadro-charmonium pentaquarks are shown in Fig. \ref{fig:octet}. Specifically, the $P_{\rm c}(4312)$, $P_{\rm c}(4440)$ and $P_{\rm c}(4457)$ are described as $N \otimes \chi_{\rm c0}(1P)$ and degenerate $N \otimes \psi(2S)$ states, respectively.
The above states have masses of $4311.9\pm0.7^{+6.8}_{-0.6}$ MeV, $4440.3\pm1.3^{+4.1}_{-4.7}$ MeV and $4457.3\pm0.6^{+4.1}_{-1.7}$ MeV, respectively; and their widths are $9.8\pm2.7^{+3.7}_{-4.5}$ MeV, $20.6\pm4.9^{+8.7}_{-10.1}$ MeV and $6.4\pm2.0^{+5.7}_{-1.9}$ MeV, respectively.
A clear and unambiguous assignment of the $P_{\rm cs}^0(4459)$ is difficult. However, purely on the basis of spectroscopy, it could be assigned to $\Lambda \otimes \psi(2S)$. 
The theoretical error in the hadro-charmonium model predictions, which are reported in Fig. \ref{fig:octet}, is $\mathcal O(100) \mbox{ MeV}$. See the supplemental material for further details.

A description of the multiplet structure of $P_{\rm c}$, $P_{\rm cs}$ and $P_{\rm css}$ states is also provided by the compact pentaquark model. 
See e.g. Ref. \cite{Santopinto:2016pkp}, in which the authors obtained a $J^P = \frac{3}{2}^-$ pentaquark octet characterized by energy spacings between strangeness = $0, -1, -2$ states, which are more regular and larger than those in Fig. \ref{fig:octet}. In this interpretation \cite{Santopinto:2016pkp}, it might be harder to accommodate non-strange [$P_{\rm c}(4440)$ or $P_{\rm c}(4457)$] and strange [$P_{\rm cs}^0(4459)$] states in the same multiplet, as done in Fig. \ref{fig:octet}. Thus, different quantum number assignments for the above states might be necessary.

Regarding the molecular model, the pentaquark $P_{\rm c}(4312)$, $P_{\rm c}(4440)$ and $P_{\rm c}(4457)$ resonances are interpreted as $\bar D \Sigma_{\rm c}$ and $\bar D^* \Sigma_{\rm c}$ molecules, as suggested e.g. by the predictions in Ref. \cite{Wu:2010jy}. 
They are three out of seven $S$-wave heavy antimeson-baryon molecules, which might constitute the first example of a full and complete HQSS multiplet for hadronic molecules \cite{Liu:2019tjn}.
Moreover, these antimeson-baryon states can be decomposed into ${\bf 3} \otimes {\bf 6} = {\bf 8} \oplus {\bf 10}$ flavor representations, which may lead to the emergence of octet and decuplet structures in the hidden-charm pentaquark sector.
By contrast, Ref. \cite{Burns:2021jlu} argues that the $P_{\rm c}(4457)$ might be very difficult to reconcile with a molecular interpretation. 

{\it Summary.} 
The flavor multiplet structure of the $1^+$ hidden-charm tetraquarks could be used to discriminate between the existing models of exotics and their assignments of the $Z_{\rm cs}(3985)$ and $Z_{\rm cs}(4000)$ states. 
As suggested in Ref. \cite{Maiani:2021tri} within the compact tetraquark interpretation, the recently discovered $Z_{\rm cs}$s might be assigned to two different $1^+$ SU(3)$_{\rm f}$ meson nonets, as they are characterized by very different decay widths. Thus, in the compact tetraquark picture, it is possible to accommodate all the existing $1^+$ hidden-charm exotic candidates within two different $1^{+(+)}$ and $1^{+(-)}$ multiplets, in which the neutral-non-strange members have opposite C-parity. Moreover, in this picture, the masses of the multiplet members satisfy the equal spacing rule, i.e. the  energy splitting between states with zero-one strange quarks and that between states with one-two strange quarks is the same.

By contrast, in the hadro-charmonium description, one only has a single $1^+$ meson nonet. A possible way to preserve the hadro-charmonium interpretation might be to hypothesize that one of the two $Z_{\rm cs}^\pm$ states is hadro-charmonium and that the other has a different nature, such as that of a molecule or compact tetraquark. 
It is also noteworthy that, in the hadro-charmonium model, an effective value of the pion mass (381 MeV) rather than its experimental value (137 MeV) must be used. This is a well-known issue, and regards the nature of pions (and of the other pseudo-scalar mesons) as Goldstone bosons. The smallness of the pion mass and the irregular energy spacings between strange and nonstrange components of the $0^-$ light nonet, compared with the more regular patterns manifested in the $\frac{1}{2}^+$ ground-state baryon octet, make the study of tetraquark hadro-charmonia more difficult than that of their pentaquark counterparts. As a result, the hadro-charmonium theoretical uncertainties in the meson sector are expected to be larger than those in the baryon sector.

In the molecular explanation, the hidden-charm tetraquarks will follow the equal spacing rule (because the mesons within the molecule follow this rule), although this pattern might be modified or even disrupted by binding effects, and also by the choice of the model parameters.
One also expects the $Z_{\rm c}(3900)$ and $Z_{\rm cs}(3985)$ mesons to be SU(3) flavor partners \cite{Du:2022jjv}, the former being described as a $(D^{*0} \bar D_{\rm s} + D^0 \bar D_{\rm s}^*)$ and the latter as a $D \bar D^*$ molecule.

Finally, it should be noted that the hidden-charm exotics may be linear combinations of the compact tetraquark, hadro-charmonium and molecular model-type structures and that it would therefore be crucial to understand (whenever it is possible) what the dominant component is. For this purpose, besides multiplet structure studies, calculations of the strong decay widths within the different models might be of great help.

\begin{acknowledgments}
This work was supported by the Academy of Finland, Project No. 320062, and Istituto Nazionale di Fisica Nucleare (INFN), Italy.
\end{acknowledgments}

\begin{appendix}

\section{Supplemental material}
The $Z_{\rm cs}(3985)$ was discovered by BESIII \cite{Ablikim:2020hsk}. Its pole mass and width were determined to be $3982.5^{+1.8}_{-2.6}\pm2.1$ MeV and $12.8^{+5.3}_{-4.4}\pm3.0$ MeV, respectively \cite{Ablikim:2020hsk}; its $J^P$ quantum numbers are expected to be $1^+$. The $Z_{\rm cs}(4000)$ was discovered by LHCb \cite{Aaij:2021ivw}. It has a mass of $4003\pm6^{+4}_{-14}$ MeV and a width of $131\pm15\pm26$ MeV; its $J^P$ quantum numbers are expected to be $1^+$ with high significance \cite{Aaij:2021ivw}.
The $P_{\rm cs}(4459)$ was observed in the $J/\psi \Lambda$ invariant mass distribution \cite{Aaij:2020gdg}.
This structure is consistent with a hidden-charm pentaquark endowed with strangeness, characterized by a mass of $4458.8\pm2.9^{+4.7}_{-1.1}$ MeV and a width of $17.3\pm6.5^{+8.0}_{-5.7}$ MeV \cite{Aaij:2020gdg}. Its spin is expected to be $\frac{1}{2}$ or $\frac{3}{2}$ and its parity can be either $-1$ or $+1$ \cite{Aaij:2020gdg}.

{\it $X_{\rm c}$, $Z_{\rm c}$ and $Z_{\rm cs}$ states in the hadro-charmonium model. Model parameters.} 
There is no complete hadro-charmonium study of the $1^{+}$ multiplet members.
The existing model predictions for $Z_{\rm cs}$s are taken from \cite[Table 4]{Ferretti:2020ewe}. 
The masses of the $X_{\rm c}$ and $Z_{\rm c}$ states at around 3.9 GeV, which are necessary in order to complete the $1^{+}$ flavor multiplet, can be calculated by using the same value of the charmonium diagonal chromo-electric polarizability $\alpha_{\psi\psi}(2S)$ as that in Refs. \cite{Ferretti:2020ewe,Anwar:2018bpu}.
The light meson radii considered here are $R_\pi =$ 0.474 fm, $R_\eta =$ 0.659 fm and $R_{\eta'} =$ 0.729 fm, and their masses are extracted from the PDG, except for that of the pion\footnote{It is well-known that the smallness of the pion mass can cause several computational problems. This is why calculations are sometimes performed by using an effective value instead of the experimental one. 
Here, an effective value of the pion mass, $\tilde M_\pi = M_K - (M_{K^*} - M_\rho) = 381$ MeV, is considered. 
The effective value of the pion radius that is used here, $R_\pi = 0.474$ fm, is the average between the value of the pion radius from the relativized QM \cite{Godfrey:1985xj} and the value of the kaon radius \cite{Zyla:2020zbs}.}.
One obtains: $M_{\psi(2S) \otimes \pi} = 3956$ MeV, $M_{\psi(2S) \otimes \eta} = 4099$ MeV and $M_{\psi(2S) \otimes \eta'} = 4217$ MeV. 

{\it $P_{\rm c}$, $P_{\rm cs}$ and $P_{\rm css}$ pentaquarks. Model parameters.} 
In the hadro-charmonium model, $P_{\rm c}$, $P_{\rm cs}$ and $P_{\rm css}$ pentaquarks are described as the bound states of charmonia and light baryons.
To obtain a full description of them, one can extract the predictions for charmonium $\otimes$ $N$, $\Sigma$ and $\Xi$ from \cite[Table 2]{Ferretti:2020ewe} and \cite[Table 2]{Anwar:2018bpu}. The masses of the charmonium $\otimes$ $\Lambda$ multiplet members, not given in the previous studies, can be computed by using: the same values of the charmonium diagonal chromo-electric polarizabilities as those in Refs. \cite{Ferretti:2020ewe,Anwar:2018bpu}; a value of the $\Lambda$ radius $R_{\Lambda} = R_{\Sigma}$, with $R_{\Sigma} = 0.863$ fm \cite[Table 1]{Ferretti:2020ewe}.
The most likely assignments to the lowest $J^P = \frac{1}{2}^+$, $\frac{1}{2}^-$ and $\frac{3}{2}^-$ hadro-charmonium pentaquarks are shown in Fig. 2. 

The above assignments can be re-examined in light of the recent LHCb finding of a $P_{\rm c}(4337)$ pentaquark, which is not consistent with the previously observed $P_{\rm c}$ states \cite{LHCb:2021chn}.
The $P_{\rm c}(4337)$ does not fit well into the description in Fig. 2 as the difference between the mass of $N \otimes \chi_{\rm c0}(1P)$ and the data is too large. 
A possible way to accommodate the $P_{\rm c}(4337)$ more properly in this scheme is to re-fit the $\alpha_{\psi \psi}(1P)$ parameter to its mass, while $\alpha_{\psi \psi}(2S)$ is constrained by the prescription of \cite[Eq. (15)]{Anwar:2018bpu}. One obtains: $\alpha_{\psi \psi}(1P) =  6.4$ GeV$^{-3}$ and $\alpha_{\psi \psi}(2S)_{c \bar c} = 10$ GeV$^{-3}$.
This second description, which is based on the new choice of the $\alpha_{\psi \psi}$ parameters, is unsatisfactory, as the masses of the $P_{\rm cs}$ pentaquarks are shifted $100-200$ MeV above those reported in Fig. 2. 
However, the new results can be used to estimate the $P_{\rm c}(4337)$ mass with its theoretical error, $M_{P_{\rm c}(4337)}^{\rm hc} = 4274 \pm \mathcal O(100) \mbox{ MeV}$, which is compatible with the experimental value \cite{LHCb:2021chn}. $\mathcal O(100) \mbox{ MeV }$ is also regarded as the average error in the hadro-charmonium model predictions.

\end{appendix}

\end{document}